\PassOptionsToPackage{hyphens}{url}
\PassOptionsToPackage{inline}{enumitem}
\documentclass[sigplan,nonacm]{acmart}

\usepackage[T1]{fontenc}
\usepackage[utf8]{inputenc}
\usepackage{letltxmacro}
\usepackage{todonotes}
\usepackage{booktabs}
\usepackage{amssymb}
\usepackage{amsmath}
\usepackage{xspace}
\usepackage{listings}
\usepackage{enumitem}
\usepackage{pifont}
\usepackage{hyperref}
\usepackage{multirow}
\usepackage{pbox}
\usepackage{color}
\usepackage[acronym]{glossaries}
\usepackage[olditem,oldenum]{paralist}
\usepackage{placeins}
\usepackage{subcaption}
\usepackage{adjustbox}

\graphicspath{{figures/}}

\definecolor{lst:comment}{rgb}{0.0,0.0,0.0}
\definecolor{lst:instructions}{rgb}{0.4,0.1,0.1}
\definecolor{lst:registers}{rgb}{0.1,0.4,0.1}
\definecolor{lst:bg}{rgb}{0.98,0.98,0.98}
\lstdefinelanguage{aarch64}{
    keywords={pacga,stp,mov,bl,ldp,cmp,jnz,ret,str,ldr,autia,pacia,eor,autib,pacib,b,cbnz,cbz,sub,pacda,autda,stur,movk},
    keywordstyle=\bfseries\color{lst:instructions},
    keywords=[2]{LR, CR, Xd, Xr, SP,XZR, X28},
    keywordstyle=[2]\color{lst:registers},
    breaklines=true,
    morecomment=[l]{;},
    commentstyle=\itshape\color{lst:comment},
}
\lstdefinestyle{customasm}{language=aarch64}
\lstdefinestyle{customc}{language=c,
    basicstyle=\small\ttfamily,
    keywordstyle=\bfseries\color{green!40!black},
    commentstyle=\itshape\color{purple!40!black},
    identifierstyle=\color{blue},
    stringstyle=\color{orange},
}

\newcommand{\inputAsmListing}[2]{\lstinputlisting[float=,floatplacement=tp,showstringspaces=false,style=customasm,linerange={2-100},label={lst:#1},caption={#2}]{listings/#1.s}}

\newif\ifabridged{}
\def\isabridged{1}
\ifdefined\isabridged{}\abridgedtrue{}\fi
\newif\ifnotabridged{}
\ifabridged\notabridgedfalse{}\else\notabridgedtrue{}\fi

\newif\ifanonymous{}
\ifdefined\isanonymous{}\anonymoustrue{}\fi
\newif\ifnotanonymous{}
\ifanonymous\notanonymousfalse{}\else\notanonymoustrue{}\fi

\newif\ifbw{}
\ifdefined\isbw{}\bwtrue{}\fi
\LetLtxMacro{\todonote}{\todo}
\renewcommand{\todo}[2][]
{\todonote[inline, caption={#2}, size=\footnotesize, #1]
{\renewcommand{\baselinestretch}{0.5}\selectfont#2\par}}
\newcommand{\instr}[1]{\texttt{\lowercase{#1}}\xspace}
\newcommand{\func}[1]{\texttt{#1}\xspace}

\newcommand{\theAttacker}{\ensuremath{\mathcal{A}}\xspace}

\newcommand{\pcanMod}{\ensuremath{\texttt{mod}}\xspace}

\newcommand{\flagSignReturnAddress}{\texttt{-msign-return-address}\xspace}
\newcommand{\flagStackProtector}{\texttt{-fstack-protector}\xspace}
\newcommand{\flagNoStackProtector}{\texttt{-fno-stack-protector}\xspace}
\newcommand{\flagStackProtectorAll}{\texttt{-fstack-protector-all}\xspace}
\newcommand{\flagStackProtectorStrong}{\texttt{-fstack-protector-strong}\xspace}

\newcommand{\longName}{\protect{Protecting the stack with PACed canaries}\xspace}
\newcommand{\implName}{\protect{\textsf{PCan}}\xspace}
\newcommand{\designName}{\protect{\textsf{PCan}}\xspace}

\newcommand{\material}{\todo{add \\material URL}}
\glsdisablehyper
\setlist[enumerate]{topsep=0pt,itemsep=-1ex,partopsep=1ex,parsep=1ex}
\setlist[itemize]{topsep=0pt,itemsep=-1ex,partopsep=1ex,parsep=1ex}

\makeatletter
\def\blfootnote{\xdef\@thefnmark{}\@footnotetext}
\makeatother

\ifnum\ifnotabridged1\else\ifanonymous1\else0\fi\fi=1
\renewcommand\footnotetextcopyrightpermission[1]{}
\fi

\newacronym{bti}{BTI}{Branch Target Indicator}
\newacronym{cfg}{CFG}{control-flow graph}
\newacronym{cfi}{CFI}{control-flow integrity}
\newacronym{dep}{DEP}{data-execution prevention}
\newacronym{dop}{DOP}{data-oriented programming}
\newacronym{el}{EL}{exception level}
\newacronym{fvp}{FVP}{Fixed Virtual Platform}
\newacronym{ir}{IR}{Intermediate Representation}
\newacronym{isa}{ISA}{instruction set architecture}
\newacronym{mac}{MAC}{message authentication code}
\newacronym{mmu}{MMU}{memory management unit}
\newacronym{pa}{PA}{pointer authentication}
\newacronym{pac}{PAC}{pointer authentication code}
\newacronym{rop}{ROP}{return-oriented programming}
\newacronym{sp}{SP}{stack pointer}
\newacronym{toctou}{TOCTOU}{time-of-check-time-of-use}
\newacronym{va}{VA}{virtual address}

\begin{document}

\title{\longName}

\ifnotanonymous

\author{Hans Liljestrand}
\orcid{1234-5678-9012}
\affiliation{\institution{Aalto University, Finland}\institution{Huawei Technologies Oy, Finland}}
\email{hans@liljestrand.dev}

\author{Zaheer Gauhar}
\affiliation{\institution{Aalto University, Finland}}
\email{zaheer.gauhar@aalto.fi}

\author{Thomas Nyman}
\affiliation{\institution{Aalto University, Finland}}
\email{thomas.nyman@aalto.fi}

\author{Jan-Erik Ekberg}
\affiliation{\institution{Huawei Technologies Oy, Finland}\institution{Aalto University, Finland}}
\email{jan.erik.ekberg@huawei.com}

\author{N. Asokan}
\affiliation{\institution{University of Waterloo, Canada}}
\email{asokan@acm.org}

\renewcommand{\shortauthors}{Liljestrand, et al.}

\else
\fi

\begin{abstract}

Stack canaries remain a widely deployed defense against memory corruption attacks.
Despite their practical usefulness, canaries are vulnerable to memory disclosure and brute-forcing attacks.
We propose \implName, a new approach based on ARMv8.3-A \gls{pa}, that uses dynamically-generated canaries to mitigate these weaknesses and show that it provides more fine-grained protection with minimal performance overhead.

\glsreset{pa}

\end{abstract}

\maketitle

\section{Introduction}

Run-time attacks that exploit memory errors to corrupt program memory are a prevalent threat.
Overflows of buffers allocated on the stack are one of the oldest known attack vectors~\cite{One96, Peslyak97}.
Such exploits corrupt local variables or function return addresses.
Modern attacks techniques---such as \gls{rop}~\cite{Shacham2007} and \gls{dop}\cite{Hu16}--- can use this well-known attack vector to enable attacks which are both expressive and increasingly hard to detect.
The fundamental problem is insufficient bounds checking in memory-unsafe languages such as C / C++.
Approaches for hardening memory-unsafe programs have been proposed, but tend to incur high performance overheads, and are therefore impractical to deploy~\cite{Szekeres2013}.
An exception is a technique called \emph{stack canaries}~\cite{Cowan98}, which is both efficient and can detect large classes of attacks.
Consequently stack canaries are widely supported by compilers and used in all major operating systems today~\cite{Cowan98,Litchfield03}.

Widely deployed stack canary implementations suffer from one or more of the following weaknesses: they \begin{inparaenum}[1)]
\item rely on canary value(s) that are fixed for a given run of a program~\cite{Cowan98};
\item store the reference canary in insecure memory, where an attacker can read or overwrite it~\cite{Litchfield03}; or
\item use only a single canary per stack frame and therefore cannot detect overflows that corrupt only local variables.
\end{inparaenum}

The recently introduced  ARMv8.3-A \gls{pa}~\cite{ARMv8A} hardware can be used to verify return addresses~\cite{Qualcomm17}, effectively turning the return address itself into a stack canary.
However, \gls{pa} on its own is susceptible to \emph{reuse attacks}, where an attacker substitutes one authenticated pointer with another~\cite{Liljestrand19a}. State-of-the-art schemes harden \gls{pa} return-address protection to ensure that protected return address as statistically unique to a particular control-flow path, and therefore cannot be substituted by an attacker~\cite{Liljestrand19b}.

We propose fine-grained \gls{pa}-based canaries which:
\begin{inparaenum}[1)]
\item protect individual variables from buffer overflow,
\item do not require secure storage for reference canary values,
\item can use existing return-address protection~\cite{Liljestrand19b} as an \emph{anchor} to produce canaries which are statistically unique to a particular function call, and
\item are efficient, since they can leverage hardware \gls{pa} instructions both for canary generation and verification.
\end{inparaenum}

Our contributions are:
\begin{itemize}
\item \textbf{\designName}: A \textbf{fine-grained}, and \textbf{efficient} \gls{pa}-based canary scheme (Section~\ref{sec:design}).
\item A \textbf{realization of \designName} for LLVM (Section~\ref{sec:implementation}).
\item \textbf{Evaluation} demonstrating that \designName is both more secure than existing stack canaries and has only a small performance impact on the protected application (Section~\ref{sec:evaluation}).
\end{itemize}

\section{Background}

\begin{figure}[tp]
\centering
\includegraphics[width=0.7\columnwidth]{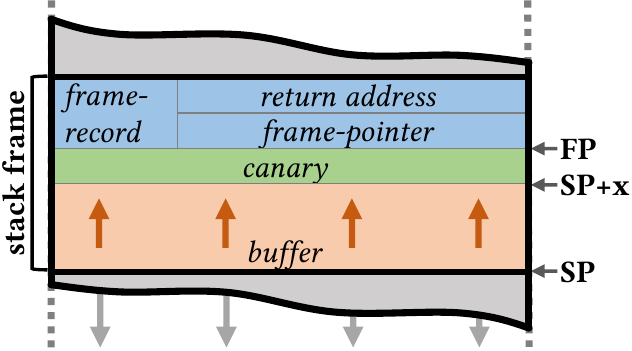}
\caption{
A stack canary is a value placed on the stack so that it will be overwritten by a stack buffer that overflows to the return address.
It allows detection of overflows by verifying the integrity of the canary before function return.
}
\label{fig:canary}
\end{figure}

A \emph{stack canary}\cite{Cowan98} is a value placed on the stack such that a stack-buffer overflow will overwrite it before corrupting the return address (Figure~\ref{fig:canary}).
The buffer overflow can then be detected by verifying the integrity of the canary before performing the return.

The initially proposed canaries were randomly generated 32-bit values assigned at process startup and stored within the process memory~\cite{Cowan98}.
The canaries must remain confidential to prevent an attacker \theAttacker from avoiding detection by writing back the correct canary when triggering the buffer overflow.
\emph{Terminator canaries}~\cite{Cowan99}, consisting of string terminator values (e.g., \texttt{0x0}, \texttt{EOF}, and \texttt{0xFF}) can prevent \theAttacker from using string operations to read or write whole canaries, thereby thwarting run-time canary harvesting.
Another approach is to re-generate canaries at run-time, for instance by masking them with the return address~\cite{Etoh00}.
However, such techniques rely on the secrecy of the return address.

Multi-threading and forked processes are another challenge; a child process or thread using the same canaries as the parent could be abused to perform a byte-by-byte guessing without relying on memory disclosure vulnerabilities.
This is particularly useful for the attacker if the child process or thread is restarted by the parent after a crash as a result of an unsuccessful attack.
In such a scenario an attacker can execute a large number of guesses without being detected.

Adversaries with arbitrary memory read or write access can trivially circumvent any canary based solution; using reads alone allows \theAttacker to first read the correct canaries from memory and then perform a sequential overwrite that writes back the correct canaries while corrupting other data.

\subsection{Stack canaries in modern compilers}

Modern compilers such as LLVM/Clang and GCC provide the \flagStackProtector feature that can detect stack-buffer overflows\footnote{https://lists.llvm.org/pipermail/cfe-dev/2017-April/053662.html}.
It is primarily designed to detect stack overflows that occur in string manipulation. The default \flagStackProtector option includes a canary only when a function defines a character array that is larger than a particular threshold. The default threshold value in GCC and LLVM is 8 characters, but in practice the threshold is often lowered to 4 to provide better coverage.
However, a stack overflow can occur on other types of variables.
The \flagStackProtectorAll option adds a canary to \textbf{all} functions.
However, it can incur a substantial use of stack space and run-time overhead in complex programs.

The \flagStackProtectorStrong option provides a better trade-off between function coverage, run-time performance, and memory cost of stack canaries. It adds a canary to any function that
\begin{inparaenum}[1)]
  \item uses a local variable's address as part of the right-hand side of an assignment or function argument,
  \item includes a local variable that is an array, regardless of the array type or length, and
  \item uses register-local variables.
\end{inparaenum}

Today \flagStackProtectorStrong is enabled by default for user-space applications in major Linux distributions, such as Debian and its derivatives\footnote{\url{https://wiki.debian.org/Hardening}}.
\flagStackProtector protects non-overflowing variables by rearranged the stack such that an overflow cannot corrupt them; but this protection cannot protect other data (e.g., other buffers).
On AArch64 the LLVM/Clang implementation of \flagStackProtector uses a single reference canary value for the whole program.
This in-memory reference canary is used to both store and verify the stack canary on function entry and return, respectively.

\subsection{ARMv8-A Pointer Authentication}
\label{sec:bg-pa}

\begin{figure}[tp]
\centering
\includegraphics[width=1\columnwidth]{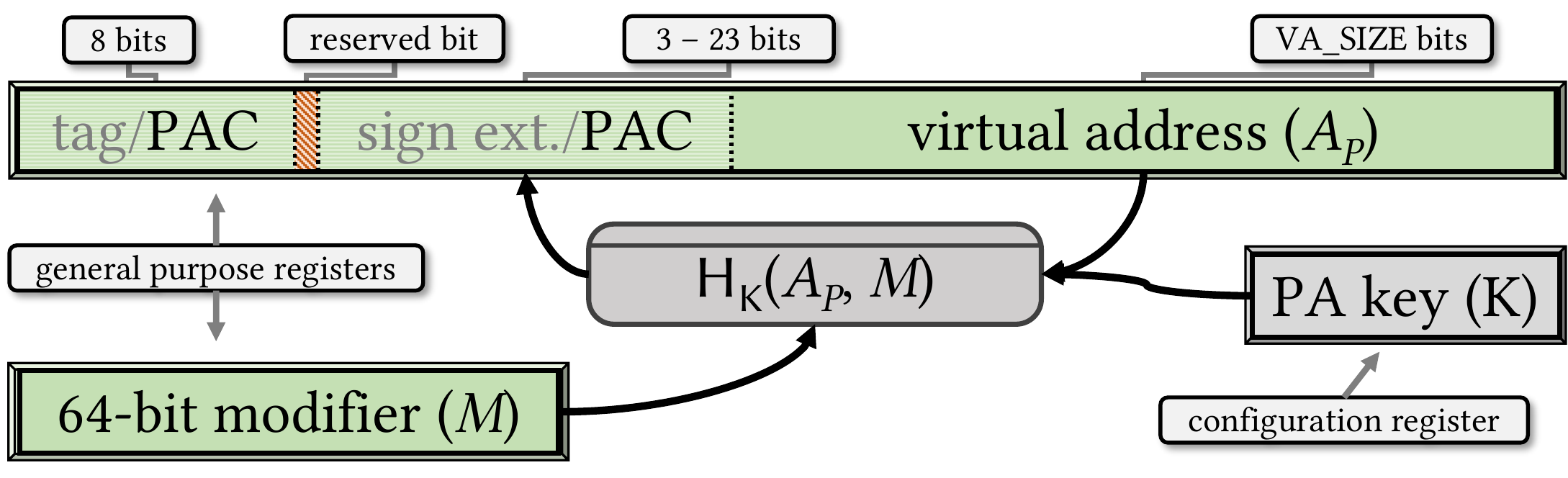}
\caption{
\Gls{pa} verifies pointers using an embedded \gls{pac} generated from a pointer's address, a 64-bit modifier and a hardware-protected key. (Figure from~\cite{Liljestrand19b})
}
\label{fig:pac}
\end{figure}

ARMv8.3-A \gls{pa} is a \gls{isa} extension that allows efficient generation and verification of \glspl{pac}; i.e., keyed \glspl{mac} based on a pointer's address and a 64-bit modifier~\cite{ARMv8M}.
The \gls{pac} is embedded in the unused bits of a pointer (Figure~\ref{fig:pac}).
On 64-bit ARM, the default Linux configuration uses 16-bit \glspl{pac}.
GNU/Linux has since 5.0 provided support for using \gls{pa} in user-space.
\Gls{pa} provides new instructions for generating and verifying \glspl{pac} in pointers, and a generic \instr{pacga} instruction for constructing a 32-bit MAC based on two 64-bit input registers.
After a \gls{pac} is added to a pointer, e.g., using the \instr{pacia} instruction, it can be verified later using the corresponding authentication instruction, in this case \instr{autia}.
A failed verification does not immediately cause an exception. Instead, \gls{pa} corrupts the pointer so that any subsequent instruction fetch or dereference based on it causes a memory translation fault.
The \instr{pacga} instruction is an exception as it outputs the produced \gls{pac} to a given destination register; verification in this case must be performed manually by comparing register values.

\begin{figure}[tp]
\centering
\includegraphics[width=0.7\columnwidth]{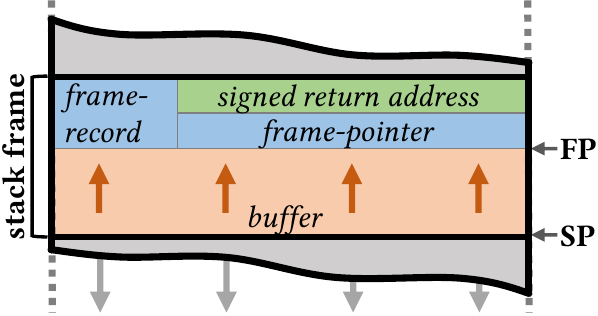}
\caption{
The signed return address generated by \flagSignReturnAddress effectively serve as a canary by allowing detection of stack-buffer overflows.
}
\label{fig:signed-return}
\end{figure}

Current versions of GCC and LLVM/Clang provide the \flagSignReturnAddress feature that uses \gls{pa} to protect return addresses~\cite{Qualcomm17}.
It signs the return address with the \gls{sp} as modifier using \instr{pacia LR, SP}.
The integrity of the return address is verified before return by issuing the corresponding authentication instruction \instr{autia LR, SP}.
Signed return addresses provides similar protection to stack canaries, i.e., if a stack-buffer overflow corrupts the return address, this is detected when the return address is verified before returning from a function (Figure~\ref{fig:signed-return}).
However, \gls{pa} is vulnerable to reuse attacks where previously encountered signed pointers can be used to used to replace latter signed pointers using the same key and modifier~\cite{Liljestrand19a}.
For instance, \flagSignReturnAddress can be circumvented by reusing a prior return address signed using the same \gls{sp} value.

\section{Adversary Model}

In this work we consider an adversary \theAttacker that attempts to compromise the memory safety of a user-space process by exploiting a stack-buffer overflow.
We do not consider adversaries at higher privilege levels, e.g., kernel level access.
However, \theAttacker can:
\begin{itemize}
\item trigger any existing stack-buffer overflow,
\item use stack-buffer over-reads to read memory and
\item repeatedly restart the process and any child processes or threads in an attempt to brute force canaries.
\end{itemize}
Adversaries with arbitrary memory read or write access cannot be thwarted with canary-based approaches and are beyond the scope of this work.

\begin{figure}[tp]
\centering
\includegraphics[width=0.7\columnwidth]{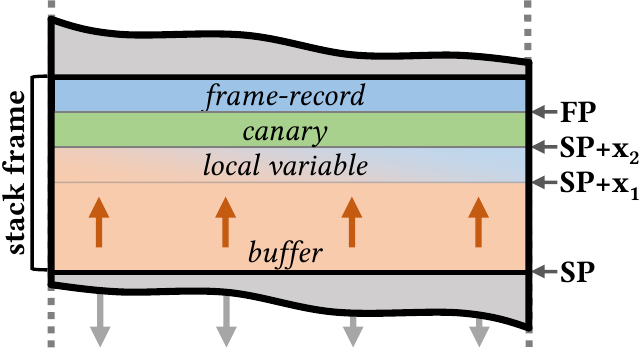}
\caption{
When using only one canary per stack frame, an attacker could overflow a stack buffer to corrupt local variables without overwriting the canary.
}
\label{fig:local-overflow}
\end{figure}

We assume that \theAttacker{} can analyze the target binary and therefore knows the exact stack layout of functions (barring dynamically allocated buffers).
This enables \theAttacker{} to target individual local variables reachable from a particular buffer overflow without overflowing canaries past the local variables (Figure~\ref{fig:local-overflow}).
If \theAttacker{} further manages to exploit a buffer over-read or other memory disclosure vulnerability, they could also overflow past the canary by simply replacing the correct canary value during the overflow.

\section{Requirements}

To detect linear buffer overflows we require a design that fulfills the following requirements:
\begin{enumerate}[label=\textbf{{R\arabic*}},ref=\textbf{{R\arabic*}}]
\item\label{req:unique} Each canary value should be statistically unique.
\item\label{req:nomod} Reference canaries must not be modifiable by an \theAttacker.
\item\label{req:overflow} A stack buffer must not be able to overflow without corrupting a canary.
\end{enumerate}

\section{Design}
\label{sec:design}

\begin{figure}[tp]
\centering
\includegraphics[width=0.7\columnwidth]{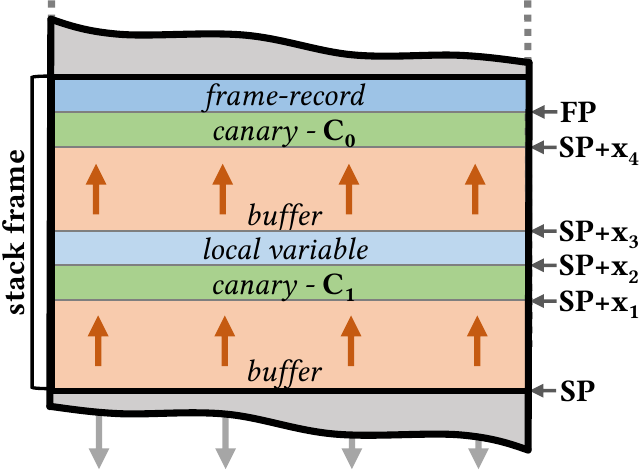}
\caption{
To detect all overflows we we inject canaries after any vulnerable stack buffer.
}
\label{fig:cauth-canary-design}
\end{figure}

We propose \designName, a \gls{pa}-based canary design that employs multiple function-specific canaries.
By placing canaries after any array that could overflow (Figure~\ref{fig:cauth-canary-design}), \designName can detect overflows that only corrupt local variables.
This prevents \theAttacker from performing precise overflows that corrupt only local variables without overwriting the canary (\ref{req:overflow}).
To exploit an overflow without detection \theAttacker is instead forced to learn the correct canary values and write them back into place.

In contrast to traditional approaches, \designName avoids attacks that leak or manipulate in-memory reference canaries by never storing reference canaries in memory (\ref{req:nomod}).
Instead, the canaries are either re-generated or verified directly using \gls{pa}.
\theAttacker thus cannot manipulate the reference canaries, and is forced to instead leak the specific on-stack canary or attempt a brute-force attack.

The canaries are generated with \gls{pa}, using a modifier consisting of a 16-bit function identifier and the least-significant 48 bits from \gls{sp}:
\[ \pcanMod = \texttt{SP}*2^{16} + \texttt{function-id} \]
This modifier makes canaries function-dependent and, when \gls{sp} differs, different for each call to the same function (\ref{req:unique}).
Because the canaries are generated at run-time and the \gls{pa} keys are randomly set on each execution, the generated canaries are also statistically unique for each execution.
To avoid detection \theAttacker must acquire the exact stack-canary belonging to the specific function call and cannot rely on pre-calculated canaries or those belonging to other function calls.

\subsection{PA-based canaries}
\label{sec:design-pa}

\Gls{pa}-based return-address protection~\cite{Qualcomm17,Liljestrand19a,Liljestrand19b} already effectively serves as a canary by detecting return-address corruption.
We therefore propose a design that can be efficiently and easily integrated with existing return-address protection schemes, but also provide a stand-alone setup.
The first canary in a function's stack frame, protecting the return-address, is either a \instr{pacga}-generated stand-alone canary or the signed return address:
\[ C_0 = \begin{cases}
    \texttt{pacga}(\texttt{SP}, \pcanMod)   & \quad \text{if stand-alone}\\
    \texttt{signed\_return\_address}              & \quad \text{if combined}\\
\end{cases}
\]
We denote a canary loaded from the stack with $C'$ to indicate that it might have been corrupted by \theAttacker.
Verification of $C'_0$ is done either by re-generating the stand-alone \instr{pacga} canary or by relying on the return-address protection to verify it.
To verify using \instr{pacga} we re-generate $C_0$ and then check that $C_0 = C'_0$.

Subsequent canaries, $C_i,i>0$, consist of signed pointers to the previous canary:
\[ C_i = \texttt{pacda}(Cptr_{i-1}, \pcanMod), i > 0 \]
where $Cptr_i$ is a pointer to $C_i$.
Verification of $C'_i,i>0$ is done by authenticating and loading the canary to retrieve $C'_{i-1}$.
If any $C'_i$ is corrupted, authentication fails, causing the subsequent load to fault (Section~\ref{sec:bg-pa})
A successful chain of loads will yield $C_0$, which is then verified as detailed above.

Our stand-alone scheme is more powerful than \flagSignReturnAddress in that it does not rely solely on the \gls{sp} value.
However, other schemes might provide better protection for the return address.
For instance, PACStack~\cite{Liljestrand19b} proposes a scheme that uses statistically unique modifiers to protect return addresses by maintaining the head of a chain of \glspl{pac} in a single reserved register.
We propose that \designName could be combined with such a mechanism by defining \pcanMod as the PACStack authentication token $auth_i$ and $C_0$ as the PACStack protected return address.
Because the \pcanMod in this case would be  statistically unique to a specific call-flow this would also harden the canaries $C_i \text{ for } i>0$.

\section{Implementation}
\label{sec:implementation}

We implement \implName as an extension to LLVM 8.0 and using the stand-alone \instr{pacga} approach (Section~\ref{sec:design-pa}).
To instrument the LLVM \gls{ir} we added new LLVM intrinsics for generating and verifying \implName canaries.
These intrinsics, along with instructions for storing and loading the canaries, are added through \gls{ir} transformations before entering the target-specific compiler backend.
We define corresponding target-specific intrinsics to leverage built-in register allocation before converting the intrinsics to hardware instructions in the pre-emit stage.

\subsection{Canary creation}

To instrument the function prologue \implName locates LLVM \func{alloca} instructions that allocate buffers in the \texttt{entry} basic block of each function.
A new 64-bit allocation for the canaries is added after each existing \texttt{alloca}.
Intrinsics for generating the canaries and storing them are then added.
The instrumented code will generate a larger stack-frame to accommodate the canaries and include code to generate and store the canary values (Listing~\ref{lst:instr-prologue}).

\inputAsmListing{instr-prologue}{
For a function with two vulnerable stack buffers \implName generates and stores two canaries.
}

\subsection{Canary verification}

To verify canaries in the function epilogue, \implName loads them in reverse order, starting from the last $C_n$ (Listing~\ref{lst:instr-epilogue}).
Each canary $C'_i$ is authenticated using \texttt{autda} and then dereferenced to acquire the next canary $C'_{i-1}$.
To verify the final canary, $C'_0$, \implName first re-generates $C_0$ and then performs a value comparison.
Upon failure, an error handler is invoked, otherwise the function is allowed to return normally.
As suggested in Section~\ref{sec:design-pa}, the final canary can be replaced with a return-address protection scheme.
The return address then serves as a canary that is verified using the corresponding protection scheme (e.g., \flagSignReturnAddress).

\inputAsmListing{instr-epilogue}{
To verify the integrity of canaries \implName first loads $C'_1$, then authenticates it before using it to load $C'_0$, which in turn is compared to the re-generated $C_0$.
}

\section{Evaluation}
\label{sec:evaluation}

Due to lack of publicly available \gls{pa}-capable hardware we have used an evaluation approach similar to prior work~\cite{Liljestrand19a,Liljestrand19b}.
We used the ARMv8-A \emph{Base Platform \gls{fvp}, based on Fast Models 11.5}, which supports ARMv8.3-A for functional evaluation.
For performance evaluation we used the PA-analogue from prior work~\cite{Liljestrand19b} and performed measurements on a 96board Kirin 620 HiKey (LeMaker version) with an ARMv8-A Cortex A53 Octa-core CPU (1.2GHz) / 2GB LPDDR3 SDRAM (800MHz) / 8GB eMMC, running the Linux kernel $\mathrm{v}4.18.0$ and BusyBox $\mathrm{v}1.29.2$.

\subsection{Performance}

We evaluated the performance of \implName using SPEC CPU 2017\footnote{\url{https://www.spec.org/cpu2017/}} benchmark package, and running it on the HiKey board.
We cross-compiled the benchmarks on an x86 system using whole program LLVM\footnote{\url{https://github.com/travitch/whole-program-llvm}}, and timed the execution of the individual benchmark programs using the \texttt{time} utility.
Results are reported normalized to a baseline measured without \implName instrumentation and compiled with \flagNoStackProtector (Figure~\ref{fig:nbench_results}).
We compare this baseline to two different setups; one using only \flagStackProtectorStrong and another using \flagNoStackProtector and \implName instrumentation.
Our results indicate that \implName incurs a very low overhead with a geometric mean of $0.30\%$.
In some cases \flagStackProtectorStrong caused the benchmarks, we suspect this is caused by it rearranging the stack.
Measurements were repeated 20 times and all binaries were compiled with \texttt{-O2} optimizations enabled.

\begin{table}
\begin{tabular}{lclcl}
benchmark      & \multicolumn{2}{c}{stack-protector} & \multicolumn{2}{c}{\implName} \\
\toprule{}
505.mcf\_r     &	$-4.78\%$ & ($4.55$) & $-0.05\%$ & ($0.13$) \\
519.lbm\_r     &	$-0.01\%$ & ($0.01$) & $ 0.04\%$ & ($0.02$) \\
525.x264\_r    &	$-0.01\%$ & ($0.01$) & $ 1.80\%$ & ($0.01$) \\
538.imagick\_r &	$-0.01\%$ & ($0.01$) & $ 0.19\%$ & ($0.01$) \\
544.nab\_r     &	 $0.05\%$ & ($0.24$) & $-0.18\%$ & ($0.16$) \\
557.xz\_r      &	 $0.00\%$ & ($0.03$) &  $0.04\%$ & ($0.06$) \\
\midrule{}
geo.mean.      &    $-0.08\%$ &          &  $0.03\%$ &          \\
\bottomrule{}
\end{tabular}
\caption{
SPEC CPU 2017 performance overhead of \implName and \flagStackProtectorStrong compared to an uninstrumented baseline (standard error for comparison is in parenthesis).
Results indicate that both schemes incur a negligible overhead (geometric mean of $0.3\%$ and $<0\%$, respectively).
}
\end{table}

\subsection{Security}

The initial \instr{pacga} canaries used by \implName provide similar security to traditional canaries.
To perform an overflow while avoiding detection \theAttacker must achieve the following goals: \begin{inparaenum}[1)]
\item find the location of canaries in relation to the overflown buffer,
\item leak the specific canary values on the stack, and
\item write back the correct canaries when performing the buffer overflow.
\end{inparaenum}
In our adversary model step 1) is trivial; \theAttacker can inspect the binary to analyze the stack layout.
Step 2) could be achieved by leaking or modifying the in-memory reference values, but because \implName generates canaries on-demand, \theAttacker is forced to leak the values from the stack (\ref{req:nomod}).
Moreover, because the canaries are statistically unique to a function and \gls{sp} value \theAttacker cannot rely on finding just any canary and substitute it with one in the overflown stack frame (\ref{req:unique}).
This limits the scope of attacks, as both the memory leak and overflow must happen within the lifetime of the attacked stack frame.
By using multiple canaries---one after each buffer---\implName can detect overflows that only touch local variables (\ref{req:overflow}).
Based on our evaluation \implName thus provides comprehensive protection with an overhead similar to currently deployed defenses.

\section{Related Work}

After the seminal article ``Smashing the Stack for fun and profit''~\cite{One96}, the notion of \emph{canaries} as a protection against buffer overflow was first introduced in StackGuard~\cite{Cowan98}, and initial GCC compiler support appeared at the same time.
StackGuard proposes to use a random canary, stored at the top of the stack (or in the thread local storage memory area), during program launch to thwart canary harvesting from the compiled code.
The threat of canary harvesting and the added protection (especially for C) provided by terminator canaries was identified shortly thereafter~\cite{Cowan99}.
The problem of canary copy and re-use was already identified by \citeauthor{Etoh00} in \citeyear{Etoh00}~\cite{Etoh00}, where the stack-frame based canary protection was augmented by masking the canary value with the function return address.
Later, \citeauthor{Strackx09}~\cite{Strackx09} argue against the futility of storing secrets in program memory, which supports using \gls{pa} to generate canaries dynamically.

Another shortcoming of canary integrity are cases when the canary mechanism is subject to brute-force attacks, e.g., in the context of process forking. \theAttacker could use the canaries in forked child processes as oracles to perform brute-force guessing of canary values.
Published solutions against this form of attack includes DynaGuard~\cite{Petsios15} and DCR~\cite{Hawkins16}.
Both solutions keep track of canary positions in the code, and re-initialize all canaries in a child process, at considerable performance overhead.
DCR optimizes the canary location tracking by chaining canaries using embedded offsets - we inherit this notion of chaining canaries from their work, although we deploy these for canary validation whereas DCR uses the mechanism for canary rewriting.
By combining the \gls{sp} in the canaries \implName provides some protection against such attacks, but comprehensive protection would require a similar approach of re-initializing canaries on fork.
Finally, the polymorphic canaries by \citeauthor{Wang18}~\cite{Wang18} optimize away the need to rewrite canaries during fork, by adding a function-specific random mask to the stack canary, which effectively removes the opportunity for systematic canary brute-forcing.

Heap protection with canaries has received much less attention than stack protection, possibly because the optimal balance between validation and performance overhead when canaries are applied to the heap remains an open problem.
The first paper on the subject was \citeauthor{Robertson03} in \citeyear{Robertson03}~\cite{Robertson03}, but a more recent mechanism --- HeapSentry by \citeauthor{Nikiforakis13}~\cite{Nikiforakis13} puts effort on the unpredictability (randomness) of the heap canaries.
HeapSentry consists of a wrapper for the allocator and a kernel module, and clocks in overheads at around $12\%$.
Pointer bounds checking schemes offer protections stronger than canaries alone, but in comparison incur significant performance overheads~\cite{Szekeres2013}.

\section{Future Work}
\label{sec:future}

Our current approach only protects stack-based variables with a static size.
Canaries for dynamic allocations cannot be verified in the prologue because they might be either out of scope or overwritten by later dynamic allocations, and are currently not used by \implName.
To prevent attacks that corrupt dynamic allocations, we propose to add instrumentation that protects dynamic allocations based on their life-time, i.e., which verifies the associated canaries immediately when the allocation goes out of scope.
The existing LLVM allocation life-time tracking could be leveraged to implement this addition without significant changes to the compiler.

We plan to refine and expand our canary approach by using compile-time analysis---i.e., the StackSafetyAnalysis~\footnote{\url{https://llvm.org/docs/StackSafetyAnalysis.html}} of LLVM---to omit instrumentation of buffers that can be statically shown to be safe.
In some cases \theAttacker could achieve their goal before function return, i.e., before the canary corruption is detected.
Such attacks could be detected earlier by utilizing the StackSafetyAnalysis to add checks after vulnerable steps during function execution, before the function epilogue.

We also plan to extend \implName instrumentation to cover heap allocations, similar to HeapSentry~\cite{Robertson03}.
Because the \gls{pa}-keys are managed by the kernel, \implName could be used for HeapSentry-like consistency checks from within the kernel, e.g., before executing system-calls.

\section{Conclusion}

Canaries are a well-established protection tool against errors occurring in programs written in memory-unsafe languages.
We present \designName, which provides hardware-assisted integrity-protection for canaries, inhibiting the most prevalent canary-circumvention techniques.
Furthermore, we propose the notion of fine-grained canaries, where canaries are not only placed to protect the return address, but can be used to identify overflows even in individual data objects.
We make available \implName, a compiler prototype, and provide real world measurements outlining the performance impact of the proposed solution variants.
Finally we point to further strategies for optimizing the use of our fine-grained canaries, as well as providing a solution path for protecting dynamic allocations.

\section{Acknowledgments}

This work was supported in part by the Intel Collaborative Research Institute for Collaborative Autonomous \& Resilient Systems (ICRI-CARS), the Academy of Finland
under grant nr. 309994 (SELIoT), and Google ASPIRE award.

{\normalsize\bibliographystyle{ACM-Reference-Format}
\bibliography{references}}

\end{document}